\documentclass[twocolumn, eqsecnum, prd, superscriptaddress]{revtex4-2}
\usepackage{amsmath,amssymb}
\usepackage{amsthm}
\usepackage{appendix}
\usepackage{comment}
\usepackage[dvipdfmx]{graphicx}
\usepackage[colorlinks]{hyperref}
\usepackage{grffile}
\usepackage{mathrsfs}
\usepackage{mathtools}
\usepackage{bm}
\usepackage{url}
\usepackage[usenames]{xcolor}
\usepackage{ulem}
\usepackage{verbatim}
\usepackage{natbib}
\newcommand{\Sub}[2]{{#1}_\textrm{#2}}

\newcommand{\pdiff}[2]{\frac{\partial {#1}}{\partial {#2}}}
\newcommand{\ve}[1]{\bm{{#1}}}
\newcommand{\SupSub}[3]{{#1}^\textrm{#2}_\textrm{#3}}

\newcommand{\etal}{\textit{et al.}}

\begin{document}
\title{Deep learning for intermittent gravitational wave signals}
\author{Takahiro S. Yamamoto}
\email{yamamoto.takahiro.u6@f.mail.nagoya-u.ac.jp}
\affiliation{Department of Physics, Nagoya University, Nagoya, 464-8602, Japan}
\author{Sachiko Kuroyanagi}
\affiliation{Instituto de F\'isica Te\'orica UAM-CSIC, Universidad Auton\'oma de Madrid, Cantoblanco, 28049 Madrid, Spain}
\affiliation{Department of Physics, Nagoya University, Nagoya, 464-8602, Japan}
\author{Guo-Chin Liu}
\affiliation{Department of Physics, Tamkang University, Tamsui, New Taipei City 25137, Taiwan}
\date{\today}
\begin{abstract}
    The ensemble of unresolved compact binary coalescences is a promising source of the stochastic gravitational wave (GW) background. For stellar-mass black hole binaries, the astrophysical stochastic GW background is expected to exhibit non-Gaussianity due to their intermittent features. We investigate the application of deep learning to detect such non-Gaussian stochastic GW background and demonstrate it with the toy model employed in Drasco \& Flanagan (2003), in which each burst is described by a single peak concentrated at a time bin.
    For the detection problem, we compare three neural networks with different structures: a shallower convolutional neural network (CNN), a deeper CNN, and a residual network. 
    We show that the residual network can achieve comparable sensitivity as the conventional non-Gaussian statistic for signals with the astrophysical duty cycle of $\log_{10}\xi \in [-3,-1]$.
    Furthermore, we apply deep learning for parameter estimation with two approaches, in which the neural network (1) directly provides the duty cycle and the signal-to-noise ratio (SNR) and (2) classifies the data into four classes depending on the duty cycle value.
    This is the first step of a deep learning application for detecting a non-Gaussian stochastic GW background and extracting information on the astrophysical duty cycle.
\end{abstract}
\maketitle

\section{Introduction}
The astrophysical stochastic gravitational-wave (GW) background is one of the most interesting targets for current and future GW experiments. It originates from the ensemble of many unresolved GW sources at high redshift and contains information about the mass function and redshift distribution of the sources. 

Observations of binary black holes (BBH) and binary neutron stars (BNS) indicate that distant merger events could be detected as a stochastic GW background by the near-future ground-based detector network~\cite{LIGOScientific:2016fpe,LIGOScientific:2017zlf,KAGRA:2021kbb}. An estimation from the merger rate shows that the energy density of the background spectra of BBH- and BNS-origins would be similar, but the statistical behavior could be very different~\cite{LIGOScientific:2017zlf}. While sub-threshold BNS events typically overlap and create an approximately continuous background, the time interval between BBH events is much longer than the duration of the individual signal, and they do not overlap. Because of this, the BBH background could be highly non-stationary and non-Gaussian (sometimes referred to as intermittent or popcorn signal). The information on the non-Gaussianity could be used to disentangle the different origins of the GW sources~\cite{Braglia:2022icu}.

Detection strategies for such non-Gaussian backgrounds have been discussed in the literature. First, Drasco \& Flanagan~\cite{Drasco:2002yd} (DF03) derived the maximum likelihood detection statistic for the case of co-located, co-aligned interferometers characterized by stationary, Gaussian white noise with the burst-like non-Gaussian signals. Although the computational cost is significantly larger than the standard cross-correlation method, it has been shown that the maximum likelihood method can outperform the standard cross-correlation search. 
Subsequently, Thrane~\cite{Thrane:2013kb} presented a method that can be applied in the more realistic case of spatially separated interferometers with colored, non-Gaussian noise. Martellini \& Regimbau~\cite{Martellini:2014xia,Martellini:2015mfr} proposed semiparametric maximum likelihood estimators. While they are framed in the context of frequentist statistics, Cornish \& Romano~\cite{Cornish:2015pda} discussed the use of Bayesian model selection. Alternative methods were also discussed. Seto~\cite{Seto:2008xr,Seto:2009ju} presented the use of the fourth-order correlation between four detectors. 
Smith \& Thrane~\cite{Smith:2017vfk,Smith:2020lkj} proposed a method to use sub-threshold BBHs in the matched-filtering search and demonstrated a Bayesian parameter estimation. Subsequently, Biscoveanu et al.~\cite{Biscoveanu:2020gds} simulated the Bayesian parameter estimation of the primordial background (stationary, Gaussian) in the presence of an astrophysical foreground (non-stationary, non-Gaussian). Finally, Matas \& Romano~\cite{Matas:2020roi} showed that the hybrid frequentist Bayesian analysis is equivalent to a fully Bayesian approach and claimed that their method can be extended to non-stationary GW background. See also Ref.~\cite{Romano:2016dpx} for a comprehensive review. 

In the general context of GW data analysis, the application of deep learning has been actively studied in the last five years. George \& Huerta~\cite{George:2016hay,George:2017pmj} showed that deep neural networks can achieve a sensitivity comparable to the matched filtering for detection and parameter estimation of BBH mergers. Although their neural network does not predict the statistical error, several authors proposed a method to predict the posterior distributions~\cite{Gabbard:2019rde,Chua:2019wwt,Green:2020hst,Dax:2021tsq,Kuo:2021qtt}. Also, deep learning has been widely applied for various types of signal (e.g., BBH mergers~\cite{Gabbard:2017lja,Chatterjee:2021lit,Mishra:2022ott}, black hole ringdown GWs~\cite{Nakano:2018vay,Shen:2019vep,Yamamoto:2020rse,Bhagwat:2021kfa}, continuous GWs~\cite{Morawski:2019awi,Dreissigacker:2019edy,Beheshtipour:2020zhb,Yamamoto:2020pus, Yamamoto:2022adl}, supernovae~\cite{Astone:2018uge}, and hyperbolic encounters~\cite{Morras:2021atg}). 

In this work, we use deep learning to analyze a non-Gaussian GW background. The great advantage of deep learning is that it is computationally cheaper than the matched-filter-based approach. It is because neural networks learn the features of the data through a training process before being applied to real data.
The data analysis of a stochastic background is usually performed by dividing the long-duration data stream ($\sim$years) into short time segments (typically $192$ seconds; see e.g.~\cite{KAGRA:2021kbb}). If we want to apply the non-Gaussian statistic of DF03, it will take a much longer time to analyze each segment compared to the standard cross-correlation statistic. On the other hand, in the case of deep learning, once the training is completed, it can quickly analyze each segment and repeat the same analysis for a large number of data segments with similar feature of training data. In this way, it is expected to reduce the total time for the analysis.
Another advantage is that neural networks can extract the features which are difficult to model. 
Thus, it could be applied to various types of non-Gaussian GW backgrounds even if the source waveform is not understood well.
As a first step, we employ the toy model and the detection method proposed by DF03. We train the neural network with the dataset that is generated by the toy model and assess the neural network's performance by comparing it with their detection method.

Finally, let us mention the work by Utina~\etal~\cite{Utina:2021ipo}, which has a similar purpose and developed neural network algorithms to detect the GW background from binary black hole mergers.
In~\cite{Utina:2021ipo}, the data is split into $1$ or $2$ second segments, and the neural network is trained with the injection of binary black hole events. On the other hand, our method is based on the toy model in DF03, which does not rely on a particular burst model and is designed to analyze segments with longer duration (as long as the computational power allows). In addition, we discuss the estimation of the intermittency (astrophysical duty cycle), while Utina~\etal~focused on the detection problem.

The paper is structured as follows. In Sec.~\ref{sec: signal model and statistic}, we 
describe the signal model and the non-Gaussian statistic proposed by DF03, which is demonstrated for the comparison
in the result sections. Section.~\ref{sec: neural network} is dedicated to a review of deep learning algorithms used in this paper. Then, we show the results of the detection problem in Sec.~\ref{sec: detection} and parameter estimation in Sec.~\ref{sec: parameter estimation}. Finally, we summarize our work in Sec.~\ref{sec: conclusion}.

\section{Signal model and maximum likelihood statistic}
\label{sec: signal model and statistic}

\subsection{signal model}
\label{sec: signal model}

We use a simple toy model used in DF03. The assumptions are the followings: two detectors that are co-located and co-aligned; the detector noises are white, stationary, Gaussian, and statistically independent; each astrophysical burst is represented by a sharp peak that has support only on a discretized time grid. The methodology could be easily extended to the case of spatially separated interferometers by introducing the overlap reduction function~\cite{Allen:1997ad,Thrane:2013kb}. Detector noise, in reality, is colored and highly non-Gaussian, non-stationary, and these effects should be taken into account before applying our method to the real data.
In this paper, however, we focus on presenting the basic methodology of deep learning and the comparison with the DF03's results. The assumption on the sharp peak signal is valid if the duration of the burst is shorter than the time resolution of the detector. In that case, the observed GW strain at the burst arrival time is the averaged amplitude over the time interval between the sampled time step. However, this assumption cannot be applied to the expected astrophysical backgrounds from BBHs and BNSs, and again, we leave it as future work.

\begin{figure}[t]
    \centering
    \includegraphics[width=8cm]{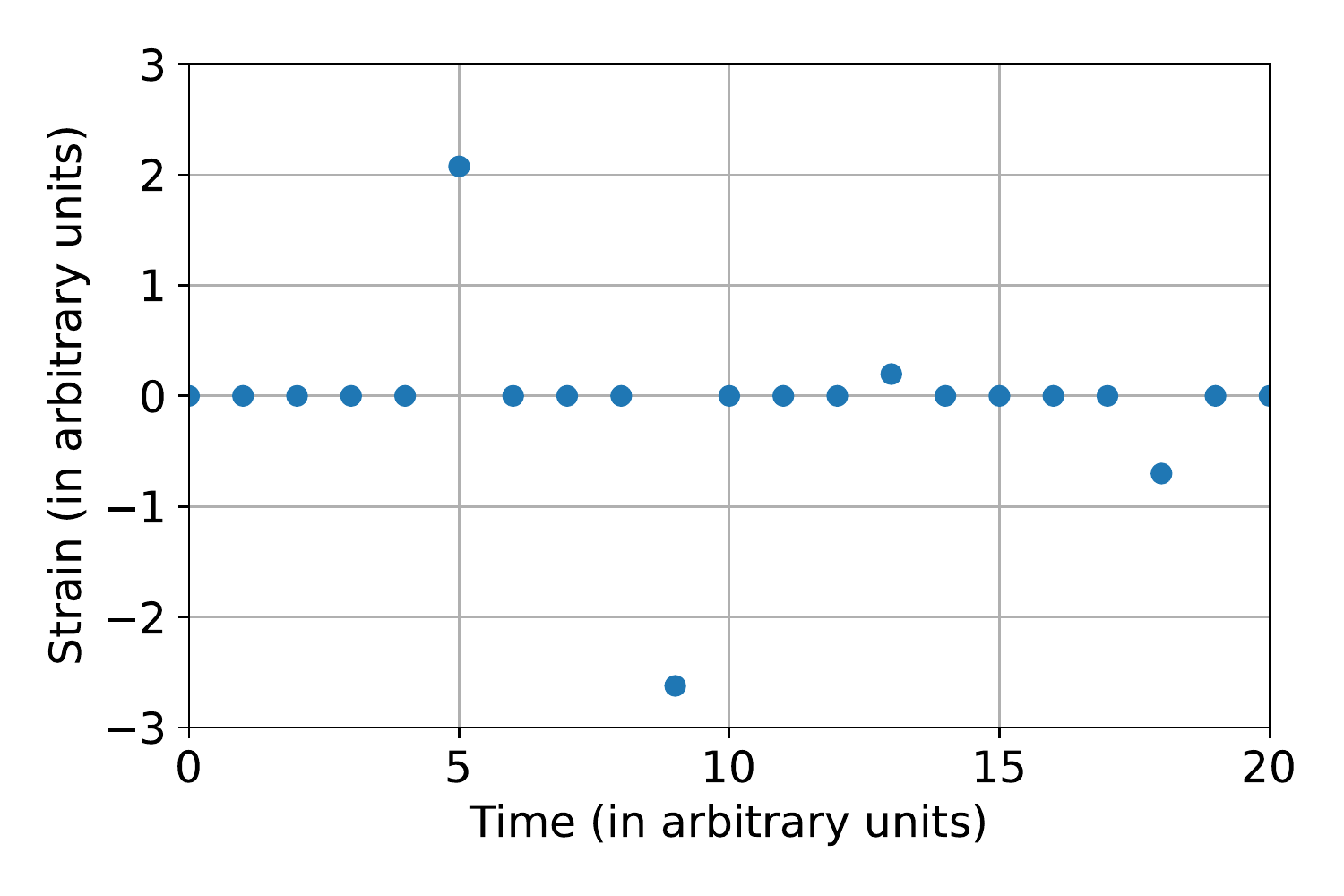}
    \caption{Example of the signal model. We see four bursts at times 5, 9, 13, and 18. Each burst is represented by a single peak.}
    \label{fig: signal example}
\end{figure}

A strain data obtained by each detector is denoted by $h^k_i$, where $i = 1,2$ labels the different
detectors, and $k = 1, 2, \cdots, N$ is a time index. We use $s^k$ to denote the GW signal. Including detector noise data, which is denoted by $n^k_i$, we can express the strain data of the $i$th detector as
\begin{equation}
    h_i^k = s^k + n_i^k\,.
    \label{eq: h=s+n}
\end{equation}
The detector noise is randomly generated from Gaussian distribution, that is,
\begin{equation}
    p(n_i^k) = \mathcal{N}(n_i^k; 0, \sigma^2_i)\,.
    \label{eq: noise model}
\end{equation}
$\mathcal{N}(x; \mu, \sigma^2)$ is a one-dimensional Gaussian distribution with a mean $\mu$ and a variance $\sigma^2$, i.e.,
\begin{equation}
    \mathcal{N}(x; \mu,\sigma^2) = \frac{1}{\sqrt{2\pi\sigma^2}} \exp\left[ - \frac{(x-\mu)^2}{2\sigma^2}\right]\,.
\end{equation}
Due to the assumption of stationarity and white, the variance $\sigma^2$ is constant in time. 
We also assume that two detectors have noise with the same variance, and we set
\begin{equation}
    \sigma^2_1 = \sigma^2_2 = 1\,,
\end{equation}
throughout this paper.

Assuming that the GW burst rate is not so high that the bursts do not overlap, we model the probability density function of the strain value at time $k$ as
\begin{equation}
    p(s^k) = \xi \mathcal{N}(s^k; 0, \alpha^2) + (1-\xi) \delta(s^k)\,,
    \label{eq: signal model}
\end{equation}
where $\alpha^2$ is the variance of the amplitude of the bursts, and $\xi$ is so-called the (astrophysical) duty cycle. The duty cycle describes the probability that a burst is present in the detector at any chosen time and takes a value in the range of $0 < \xi \leq 1$.
The case of $\xi=1$ is equivalent to Gaussian stochastic GWs. On the other hand, it is reduced to the absence of the signal for $\xi=0$. A signal exhibits non-Gaussianity as $\xi$ decreases. Figure~\ref{fig: signal example} shows an example of 
the time-series signal generated based on Eq.~\eqref{eq: signal model}.
Following DF03, we define the signal-to-noise ratio (SNR) of the non-Gaussian stochastic background by
\begin{equation}
    \rho = \frac{\xi \alpha^2 \sqrt{N}}{\sigma_1 \sigma_2}\,,
\end{equation}
and use it for describing the strength of the signal.

\subsection{Non-Gaussian statistic}
\label{sec: non Gaussian statistic}

As proposed in DF03, the likelihood ratio can be used as a detection statistic. Under the assumption of the noise model Eq.~\eqref{eq: noise model} and the signal model Eq.~\eqref{eq: signal model}, the likelihood ratio can be reduced to
\begin{equation}
    \SupSub{\Lambda}{NG}{ML} = \max_{0 <\xi \leq 1}\ \max_{\alpha^2>0}\ \max_{\sigma_1^2 \geq 0}\ \max_{\sigma_2^2 \geq 0}\ \SupSub{\lambda}{NG}{ML} (\alpha^2, \xi, \sigma_1^2, \sigma_2^2)\,,
    \label{eq: def nonGaussianLambda}
\end{equation}
where
\begin{widetext}
\begin{align}
    \SupSub{\lambda}{NG}{ML}(\alpha^2, \xi, \sigma_1^2, \sigma_2^2) &\coloneqq \prod_{k=1}^N \left\{ \frac{\bar{\sigma}_1 \bar{\sigma}_2 \xi}{\sqrt{\sigma_1^2 \sigma_2^2 + \sigma_1^2 \alpha^2 + \sigma_2^2 \alpha^2}} \exp \left[ \frac{(h_1^k/\sigma_1^2 + h_2^k/\sigma_2^2)^2 \alpha^2}{2(\alpha^2/\sigma_1^2 + \alpha^2/\sigma_2^2 + 1)} - \frac{(h_1^k)^2}{2\sigma_1^2} - \frac{(h_2^k)^2}{2\sigma_2^2} + 1 \right] \right. \notag \\
    &\ \left. + \frac{\bar{\sigma}_1 \bar{\sigma}_2}{\sigma_1 \sigma_2} (1-\xi) \exp \left[ - \frac{(h_1^k)^2}{2\sigma_1^2} - \frac{(h_2^k)^2}{2\sigma_2^2} + 1 \right]  \right\}\,,
\end{align}
\end{widetext}
and
\begin{equation}
    \bar{\sigma}^2_i \coloneqq \frac{1}{N} \sum_{k=1}^N (h_i^k)^2\,.
\end{equation}
More details of the non-Gaussian statistic are described in Appendix~\ref{appendix:NG}.

In the later section, we compare the results of the non-Gaussian statistic and the neural networks for the detection problem. As seen in Eq.~\eqref{eq: def nonGaussianLambda}, we need to perform the parameter space search to find the maximum value of $\SupSub{\lambda}{NG}{ML}$ in the four-dimensional space. To do that, we take the grid points spanning over the ranges of $\rho \in [0.0, 4.0]$, $\log_{10}\xi \in [-5.0, 0.0]$, and $\sigma_1^2, \sigma_2^2 \in [0.95, 1.05]$ with the regular interval of $\Delta\rho=0.1$, $\Delta \log_{10}\xi=0.05$, and $\Delta\sigma_1^2 = \Delta\sigma_2^2 = 0.05$.


\section{Basics of neural network}
\label{sec: neural network}

\subsection{Structure}

The fundamental unit of a neural network is called a(n) (artificial) neuron which is an artificial model of a nerve cell in a brain. 
A neuron takes values signaled by other neurons as inputs and returns a single value as an output. The alignment of neurons is called a layer, and a neural network consists of a sequence of layers. Each layer takes the output of the previous layer and passes its own output to the next layer. The input data of a neural network go through many layers, and a neural network returns the output data. In the following, we denote an input vector and an output vector of each layer by $\ve{x}$ and $\ve{y}$, respectively. The dimensions of the input and the output depend on the type of layer, which is described below.

A fully connected layer is one of the fundamental layers of a neural network. It takes a one-dimensional real-valued vector as an input and returns a linear transformation of the input data. The operation can be described as
\begin{equation}
  y_i = \sum_{j=0}^{N} w_{ij} x_j\,,
  \label{eq: fully-connected layer}
\end{equation}
where $N$ is the number of elements of the input vector. The 0th component is set to be $x_{0}=1$ and represents the constant term of the linear transformation. The coefficients $w_{ij}$ are called weight, and we must appropriately tune them before applying the neural network to real data.

A linear transformation like a fully connected layer is usually followed by a nonlinear function which is called an activation function. Most 
activation functions have no tunable parameters. In this work, we used a rectified linear unit (ReLU) defined by
\begin{equation}
    \mathrm{ReLU}(z) \coloneqq \begin{cases}
    z & \text{if } 0 \leq z \,.\\
    0 & \text{if } z < 0 \,.\\
    \end{cases}
    \label{eq: relu}
\end{equation}
An activation function can take input with arbitrary size and be applied elementwise.

For image recognition, it is important to capture the local pattern. To do so, filters with a much smaller size than that of the input data are used. A convolutional layer carries out a convolution between input data and filters. A neural network containing one or more convolutional layers is often called a convolutional neural network (CNN).
In this work, we use a one-dimensional convolutional layer. It can take a two-dimensional tensor as an input that is denoted by $\ve{x} = x_{c,i}$. This represents the situation where each grid of the data has multiple values. The different values contained in one pixel are called channels, which are represented by the first index $c$.
For example, a color image has three channels corresponding to the primary colors, namely, red, blue, and green. In our case, the strain data has two channels corresponding to two different detectors. 
The second index $i$ corresponds to a pixel.
Formally, we can write a one-dimensional convolutional layer by
\begin{equation}
  y_{c,i} = \sum_{c'=1}^{C} \sum_{k=0}^{K-1} w_{c',c,k} x_{c', s(i-1)+k}\,,
  \label{eq: convolutional layer}
\end{equation}
where the parameter $w_{c',c,k}$ characterize the filter, $K$ is the filter size, and $C$ is the number of channels. The filter is applied multiple times to the input data by sliding it over the whole matrix. The parameter $s$ is called stride and controls the step width of the slide.

A pooling layer reduces the size of data by contracting several data points into one data point. There are several variants of pooling layers depending on how to reduce information. In the present work, we use two types of pooling. The max pooling layer is defined by
\begin{equation}
  y_{c,i} = \max_{j=0,1,\dots,K-1}\left[ x_{c,s(i-1)+j} \right]\,.
  \label{eq: maxpooling layer}
\end{equation}
Also, we use the average pooling that is defined by,
\begin{equation}
    y_{c,i} = \frac{1}{K} \sum_{j=0}^{K-1} x_{c,s(i-1)+j}\,.
    \label{eq: averagepooling layer}
\end{equation}

The last layer of a neural network is called the output layer. 
It should be tailored depending on the problem to solve.
For the regression, the identity function
\begin{equation}
    y_i = x_i\,,
    \label{eq: identity output layer}
\end{equation}
is often applied. Usually, the identity function is not explicitly applied because it is trivial. On the other hand, the classification problem requires a more tricky layer. In the classification problem, the neural network is constructed in a way that each element of the output corresponds to the probability that the input is likely to belong to each class. To interpret the output as the probabilities, they must satisfy the following relations,
\begin{equation}
    \sum_{i=1}^{N_\mathrm{class}} y_i = 1\,,
    \label{eq:sum p=1}
\end{equation}
and
\begin{equation}
    y_i \geq 0 \text{ for any }i\,.
    \label{eq: p>0}
\end{equation}
Here, $N_\mathrm{class}$ is the number of target classes. A softmax layer that is defined by
\begin{equation}
  y_i = \frac{\exp[x_i] }{\sum_{j=1}^{N_\mathrm{class}} \exp[x_j]}\,,
  \label{eq: softmax layer}
\end{equation}
is suitable for the classification problem. The output of a softmax layer~\eqref{eq: softmax layer} satisfies the conditions Eqs.~\eqref{eq:sum p=1} and ~\eqref{eq: p>0}.

\subsection{Residual block}
\label{subsec: resnet}

\begin{figure}[t]
    \centering
    \includegraphics[width=7.5cm]{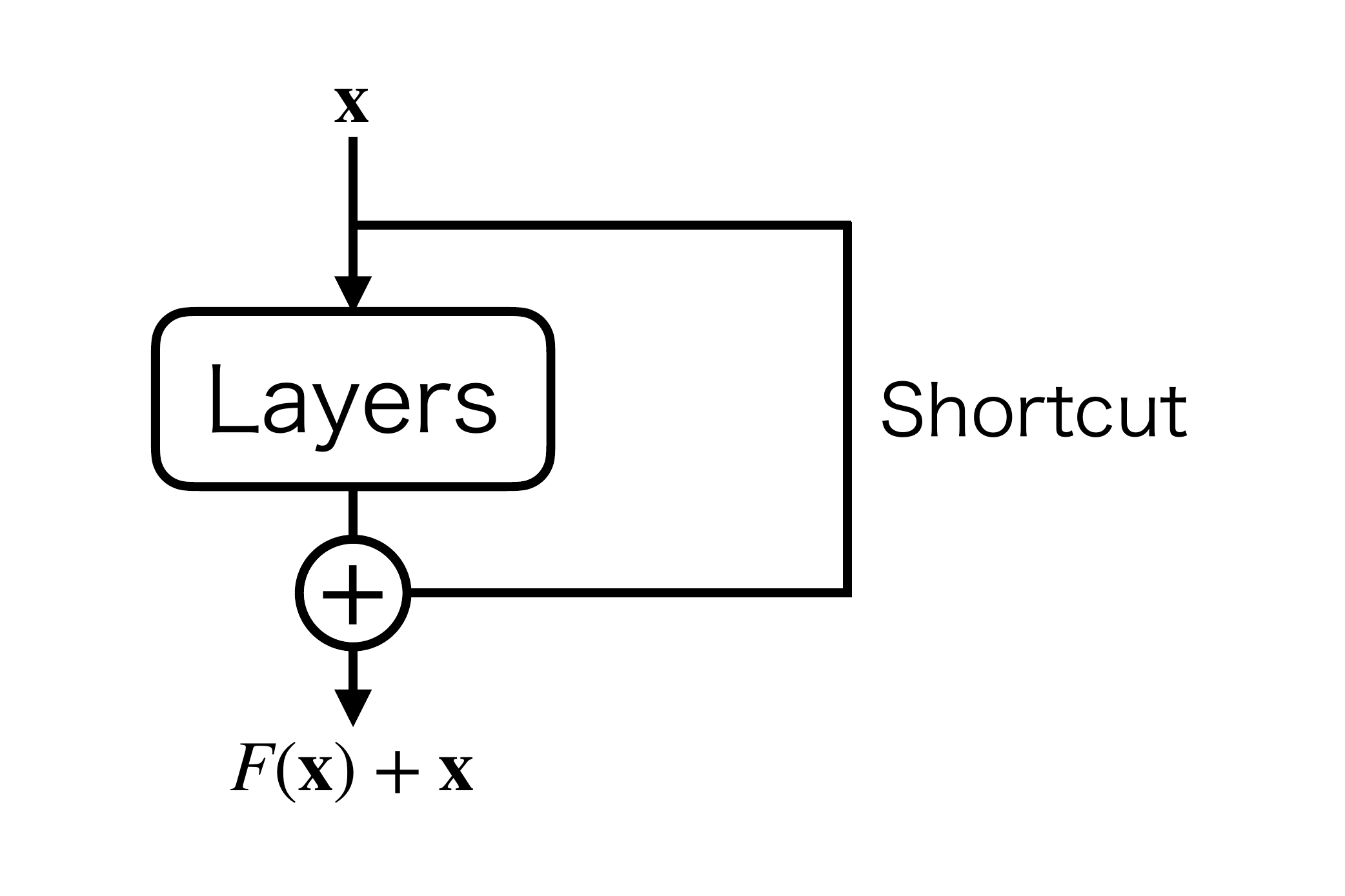}
    \caption{Schematic picture of a residual block. A standard layer transforms an input $\ve{x}$ into an output $F(\ve{x})$, while a shortcut connection directly passes the input to the output. In total, the residual block returns their sum $F(\ve{x})+\ve{x}$. 
    If the input $\ve{x}$ and the output $F(\ve{x})$ have different shapes, the data passing through the shortcut connection is reshaped appropriately to have the same shape.
    }
    \label{fig:resnet structure}
\end{figure}

One may naively expect that a deeper neural network shows better performance. This expectation is valid to some extent. However, it is empirically known that the performance gets saturated as the network depth increases. On the contrary, the accuracy gets worse. It is known as the degradation problem. He \textit{et al.}~\cite{He:2015wrn} proposed the residual learning to address the degradation problem. 
The idea of the residual network is to introduce a shortcut connection, as shown in Fig.~\ref{fig:resnet structure}, which enables us to efficiently train deep neural networks.

\begin{figure}[t]
    \centering
    \includegraphics[width=8cm]{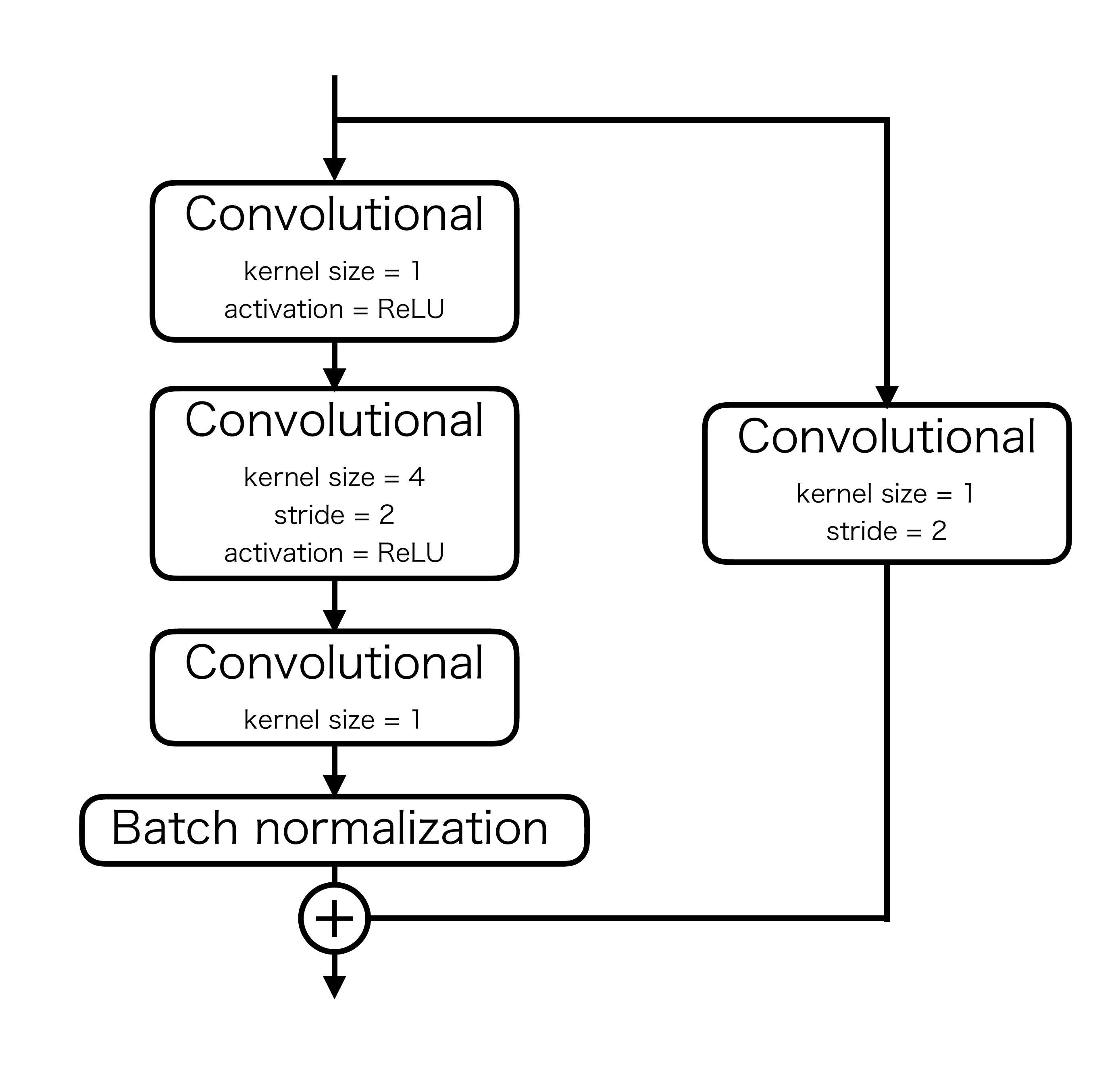}
    \caption{Structure of the residual block we used in this work. The main path consists of the three convolutional layers and the batch normalization layer. In the shortcut connection, the convolutional layer reshapes the data so that the size of the output matches that of the output of the main path.}
    \label{fig: our residual block}
\end{figure}
Figure~\ref{fig: our residual block} shows another type of residual block called a bottleneck block~\cite{He:2015wrn}. The main path has three convolutional layers. The first convolutional layer has a kernel size of one and reduces the number of channels. The second convolutional layer plays a role as a usual one. The third convolutional layer has a kernel size of 1 and recovers the number of channels. Both the first and the second convolutional layers are followed by ReLU activation~\eqref{eq: relu}. 
The batch normalization~\cite{Ioffe:2015ovl} is located at the end of the main path, where the average and the variance of the input data are calculated elementwise over the batch,
\begin{align}
    \ve{\mu} &\coloneqq \frac{1}{\Sub{N}{batch}} \sum_{n=1}^{\Sub{N}{batch}} \ve{x}_n\,, \label{eq: BN average}\\
    \ve{v} &\coloneqq \frac{1}{\Sub{N}{batch}} \sum_{n=1}^{\Sub{N}{batch}} (\ve{x}_n - \ve{\mu})^2 \label{eq: BN variance}\,.
\end{align}
Here, a batch is a subset of the training data, and $N_\mathrm{batch}$ is the size of a batch. The use of a batch in the training is explained in the later subsection, Sec.~\ref{sec: training process}.
Using Eqs.~\eqref{eq: BN average} and~\eqref{eq: BN variance}, the batch normalization transforms the input data into
\begin{align}
    \hat{\ve{x}}_n &\coloneqq \frac{\ve{x}_n - \ve{\mu}}{\sqrt{\ve{v}+\epsilon}} \label{eq: BN normalize}\,,\\
    \ve{y} &= \ve{\gamma} \ast \hat{\ve{x}}_n + \ve{\beta} \label{eq: BN output}\,,
\end{align}
where $\ve{\gamma}$ and $\ve{\beta}$ are trainable parameters, the asterisk $\ast$ represents an elementwise multiplication, and $\epsilon$ is introduced to prevent the overflow. We set $\epsilon=10^{-5}$.

The shortcut connection also has a convolutional layer with a kernel size of one and a stride of two. The input data is reshaped to match the size of the output of the main path.

\subsection{Supervised learning}

Before applying the neural network to real data, we optimize the neural network's weights using a dataset prepared in advance. 
The optimization process is called training. To train a neural network, we prepare a dataset consisting of many pairs of input data and target data, which is hereafter denoted by $\ve{t}$.
In this paper, the input data is the time-series strain data, and the target data should be chosen appropriately depending on the problem to solve.

In this work, we treat two types of problems: regression and classification. For the regression problem, the injected values of the parameters can be used as the target values. For the classification problem in which the inputs are classified into several classes, the one-hot representation is widely used for the target vector.
When the number of the target classes is $N_\mathrm{class}$, the target vector is a $N_\mathrm{class}$-dimensional vector that takes 0 or 1 for each element. If the input is assigned to the $i$th class, only $i$th element takes 1, and others are 0. For example, in the detection problem presented in Sec.~\ref{sec: detection}, we have two classes, that is, the absence and the presence of the GW signal. In this case, the target vector is chosen as
\begin{equation}
    \ve{t} = \begin{cases}
    (1,0) & \text{in the absence of a GW signal} \\
    (0,1) & \text{in the presence of a GW signal}\,.
    \end{cases}
\end{equation}
In Sec.~\ref{sec: parameter estimation}, we demonstrate estimation of the duty cycle. We first apply the ordinary parameter estimation approach; the injected values of the parameters (the signal amplitude and the duty cycle) are used as the target values. In the second approach, we reduce the parameter estimation to the classification problem, in which the inputs are classified into four classes of duty cycle values.

\subsection{Training process}
\label{sec: training process}
In the training process, we use a loss function to quantify the deviation between the neural network predictions and the target values. 
For the regression, various choice exists. In this work, we employ the L1 loss defined by
\begin{equation}
    L_\mathrm{L1}(\ve{y}, \ve{t}) = \frac{1}{N_\mathrm{param}} \sum_{i=1}^{N_\mathrm{param}} |y_i - t_i|\,,
    \label{eq:L1loss}
\end{equation}
where $N_\mathrm{param}$ is the number of parameters to be estimated.
For the classification problem, a cross-entropy loss, defined by,
\begin{equation}
  \Sub{L}{cross-entropy}(\ve{y}, \ve{t}) = - \sum_{i=1}^{N_\mathrm{class}} t_i \ln y_i\,,
  \label{eq: cross entropy discrete prob}
\end{equation}
is widely used.

The weights of a neural network are updated so that the sum of the loss functions for all training data is small.
However, in general, the minimization of the loss function cannot be done analytically. Thus, the iterative process is employed. We divide the training data into several subsets, called a batch. 
In each step of the iteration, the prediction and the loss evaluation are made for all data contained in a given batch. 
The update process depends on the gradient of the sum of the loss function over the batch, i.e.,
\begin{equation}
  \mathcal{L} = \frac{1}{\Sub{N}{batch}} \sum_{n=1}^{\Sub{N}{batch}} L(\ve{y}_n, \ve{t}_n)\,,
\end{equation}
where $\ve{y}_n$ and $\ve{t}_n$ are the prediction of the neural network and the target vector for the $n$ th data, respectively.

The simplest update procedure is the stochastic gradient descent method (SGD). In SGD, the derivatives of the loss function with respect to the neural network's weights are calculated, and the weights are updated by the procedure
\begin{equation}
    w \to w - \eta\pdiff{\mathcal{L}}{w}\,,
\end{equation}
where we omit all subscripts and superscripts of $w$ just for brevity. $\eta$ is called learning rate and characterizes how much the update of weights is sensitive to the loss function gradients. A batch is randomly chosen for every iteration step. Many updated procedures have been proposed so far. Most of them commonly exploit the gradients of the loss function with respect to the weights. In spite of the tremendous number of the weights, the algorithm called back propagation enables us to efficiently calculate all gradients of the loss function.


\section{Detection of non-Gaussian stochastic GWs}
\label{sec: detection}
In this section, we present the application of deep learning to the detection problem of the non-Gaussian stochastic GW background.

\subsection{Setup of neural networks}
\label{subsec: setup of detection NN}

\begin{table}[t]
    \caption{\label{tab: structure of shallower CNN}
    Structure of the shallower CNN. The total number of tunable parameters is 668658. The first column shows the name of the layer. The second column is the size of the output data. The last column is the number of the tunable parameters. The network first has an input layer that passes the input data, and the size of the output is equal to that of the input data. Before the flattening layer, the output size has two dimensions corresponding to the number of channels and the data length.
    The flattering layer transforms data into a one-dimensional vector. It is followed by three fully connected layers and two activation layers. The last layer is the softmax layer returning the probabilities of the absence and the presence of the signal.}
    \begin{ruledtabular}
    \begin{tabular}{lll}
        Layer & Output size & No. of parameters \\ \hline
        Input & (2, 10000) & - \\
        1D convolutional & (16, 9993) & 272 \\
        ReLU & (16, 9993) & - \\
        1D max pooling & (16, 2498) & - \\
        1D convolutional & (32, 2491) & 4128 \\
        ReLU & (32, 2491) & - \\
        1D max pooling & (32, 622) & - \\
        1D convolutional & (64, 619) & 8256 \\
        ReLU & (64, 619) & - \\
        1D max pooling & (64, 154) & - \\
        1D convolutional & (128, 151) & 32896 \\
        ReLU & (128, 151) & - \\
        1D max pooling & (128, 37) & - \\
        Flattening & (4736,) & - \\
        Fully connected & (128,) & 606336 \\
        ReLU & (128,) & - \\
        Fully connected & (128,) & 16512 \\
        ReLU & (128,) & - \\
        Fully connected & (2,) & 258 \\
        Softmax & (2,) & -
    \end{tabular}
    \end{ruledtabular}
\end{table}
\begin{table}[t]
    \caption{\label{tab: structure of deeper CNN}
    Structure of the deeper CNN. The total number of tunable parameters is 10127426. The description of the table is the same as Table.~\ref{tab: structure of shallower CNN}}
    \begin{ruledtabular}
    \begin{tabular}{lll}
        Layer & Output size & No. of parameters \\ \hline
        Input & (2, 10000) & - \\
        1D convolutional & (64, 9993) & 1088 \\
        ReLU & (64, 9993) & - \\
        1D convolutional & (64, 9986) & 32832 \\
        ReLU & (64, 9986) & - \\
        1D max pooling & (64, 2496) & - \\
        1D convolutional & (64, 2489) & 32832 \\
        ReLU & (64, 2489) & - \\
        1D convolutional & (64, 2482) & 32832 \\
        ReLU & (64, 2482) & - \\
        1D max pooling & (64, 620) & - \\
        1D convolutional & (64, 613) & 32832 \\
        ReLU & (64, 613) & - \\
        1D convolutional & (64, 606) & 32832 \\
        ReLU & (64, 606) & - \\
        1D max pooling & (64, 303) & - \\
        Flattening & (19392,) & - \\
        Fully connected & (512,) & 9929216 \\
        ReLU & (512,) & - \\
        Fully connected & (64,) & 32832 \\
        ReLU & (64,) & - \\
        Fully connected & (2,) & 130 \\
        Softmax & (2,) & -
    \end{tabular}
    \end{ruledtabular}
\end{table}

\begin{table}[t]
    \caption{\label{tab: structure of resnet}
    Structure of the residual network. Each residual block has the structure shown in Fig.~\ref{fig: our residual block}. The total number of the trainable parameters is 10280546, which is comparable to that of the deeper CNN presented in Table.~\ref{tab: structure of deeper CNN}. }
    \begin{ruledtabular}
    \begin{tabular}{lll}
        Layer & Output size & No. of parameters \\ \hline
        Input & (2,10000) & - \\
        1D convolutional & (64,9993) & 1088 \\
        ReLU & (64,9993) & - \\
        Residual block & (64,4997) & 7456 \\
        ReLU & (64, 4997) & - \\
        Residual block & (64, 2499) & 7456 \\
        ReLU & (64, 2499) & - \\
        Residual block & (64, 1250) & 7456 \\
        ReLU & (64, 1250) & - \\
        1D average pooling & (64, 312) & - \\
        Flatten & (19968,) & - \\
        Fully connected & (512,) & 10224128 \\
        ReLU & (512,) & - \\
        Fully connected & (64,) & 32832 \\
        ReLU & (64,) & - \\
        Fully connected & (2,) & 130 \\
        Softmax & (2,) & - 
    \end{tabular}
    \end{ruledtabular}
\end{table}

We test two CNNs of different sizes (deeper and shallower CNNs) and the residual network. The deeper CNN has about the same amount of tunable parameters as the residual network, which is useful for making a fair comparison with the residual network, and the shallower CNN has fewer parameters.
Two CNNs have similar structures, which are shown in Table~\ref{tab: structure of shallower CNN} and~\ref{tab: structure of deeper CNN}. 
Just before the first fully connected layer, the data is reshaped into a one-dimensional vector, which is called flattening and is often regarded as a layer.
Table~\ref{tab: structure of resnet} shows the structure of the residual network.

We train the three networks in an equal manner.
For the signal and noise model applied in this paper, generating data is not computationally costly, so we can generate data for every iteration of the training. We set the batch size to 256 and divide a batch into two subsets. The first half is the data containing only noise, and another half contains the GW signal and noise. Each data has two simulated strain data $\{h^k_1, h^k_2\}$ (see Eq.~\eqref{eq: h=s+n}) that are generated by using the noise model~\eqref{eq: noise model} and the signal model~\eqref{eq: signal model}. We assign the target vector $\ve{t}=(1,0)$ and $\ve{t} = (0,1)$ for the absence and the presence of the GW signal, respectively.
The data length is set to be $N=10^4$. 
For the signal injection, the astrophysical duty cycle is sampled from a log uniform distribution on $\xi \in [10^{-3}, 10^{-1}]$.
The SNR is uniformly sampled from $[\rho_\mathrm{min}, 4.0]$ with
\begin{equation}
    \rho_\mathrm{min} = \max[0.5, 3.5 + 1.3 \log_{10}\xi]\,.
    \label{eq: SNR lower bound}
\end{equation}
The lower bound $\rho_\mathrm{min}$ is set for the following reason.
The sensitivity of the non-Gaussian statistic depends on the duty cycle, as described in Appendix~\ref{appendix:NG}.
We expect the sensitivity of the neural networks also show this trend and not significantly outperform the non-Gaussian statistic. If we use a lower bound of SNR that is constant with the duty cycle, it could happen for a larger duty cycle that we train the neural network with wrong reference data for a positive detection, which contains too small a signal to be detected, and it can result in the degradation of the neural network. Therefore, we manually give the lower bound~\eqref{eq: SNR lower bound} on SNR that is slightly below the detectable SNR of the non-Gaussian statistic.

Before inputting the data
to the neural network, we normalize them to make the mean zero and the variance unity. Thus, the normalized input is given by
\begin{equation}
    \hat{h}^k_i = \frac{h^k_i - \Sub{\mu}{h}}{\Sub{\sigma}{h}}\,,
    \label{eq: normalized strain}
\end{equation}
where
\begin{equation}
    \Sub{\mu}{h} \coloneqq \frac{1}{2N}\sum_{k=1}^N (h^k_1 + h^k_2)\,,
\end{equation}
and
\begin{equation}
    \Sub{\sigma}{h}^2 \coloneqq \frac{1}{2N} \sum_{k=1}^N \left\{ (h^k_1-\Sub{\mu}{h})^2 + (h^k_2-\Sub{\mu}{h})^2 \right\}\,.
\end{equation}

We use the cross-entropy loss~Eq.~\eqref{eq: cross entropy discrete prob} with $N_\mathrm{class}=2$.
The weight update is repeated for 100,000 iterations. We use the Adam~\cite{Kingma:2014vow} as an update method. The learning rate is set at $10^{-5}$. The code is implemented with \textsc{pytorch}~\cite{Paszke2019}, a library for deep learning.

\subsection{Result}

\begin{figure*}[t]
    \centering
    \includegraphics[width=15cm]{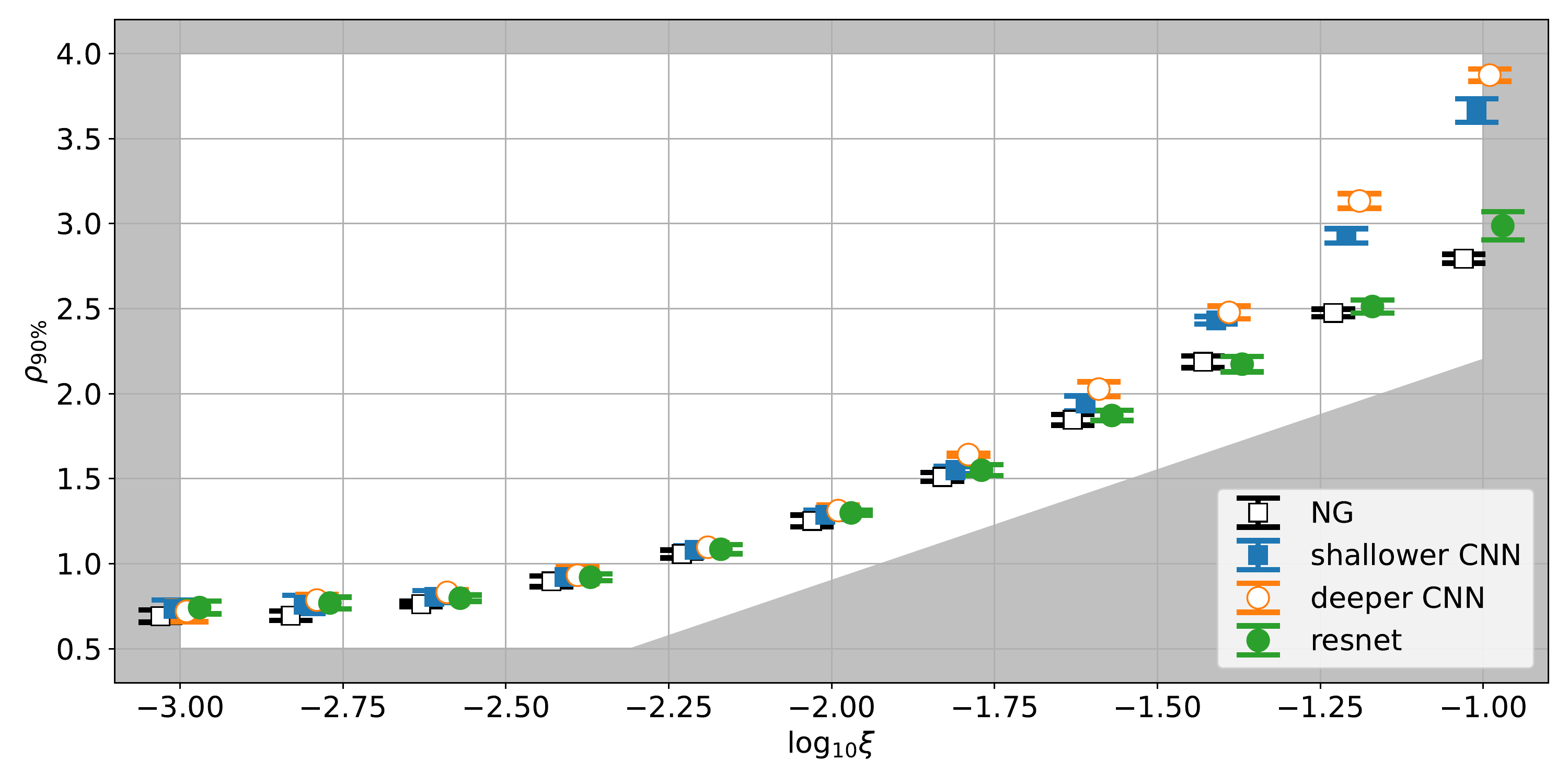}
    \caption{\label{fig:resnetresult}
    Minimum detectable SNR with 90\% detection probability for the non-Gaussian statistic, two convolutional neural networks (shallower and deeper), and the residual network. The false alarm rate is set at 5\%. The black squares are the non-Gaussian statistic, the blue squares are the shallower CNN, the orange circles are the deeper CNN, and the green circles are the residual network. For visibility, the dots are slightly shifted in the horizontal direction. The error bar shows the standard deviation of $\rho_{90\%}$ evaluated by four independent runs. The shaded area is the parameter region not used for training.}
\end{figure*}

Now we evaluate the detection efficiencies of the neural networks and compare them with the non-Gaussian statistic. First, we set the thresholds of the detection statistics by simulating noise-only data. The false alarm probability is the fraction of false positive events over the total test events, i.e.,
\begin{equation}
    \mathrm{FAP} = \frac{N(\Gamma_\ast<\Gamma)}{N_\mathrm{noise}}\,,
    \label{eq:calcFAP}
\end{equation}
where $N_\mathrm{noise}$ is the number of the simulated noise data, $\Gamma$ is the detection statistic, $\Gamma_\ast$ is the threshold value of $\Gamma$, and $N(\Gamma_\ast < \Gamma)$ is the number of events that the detection statistic exceeds the threshold. 
The neural network returns the probability of each class, denoted by $\{p_i\}_{i=1,\dots,N}$ which satisfies $\sum_{i=1}^{N}p_i=1$. Now we have the two classes ($N=2$) corresponding to the absence and presence of a GW signal.
We set $\Gamma = p_2$, which is the probability that the data contains a GW signal, for the neural networks and $\Gamma = \Lambda^\mathrm{NG}_\mathrm{ML}$ for the non-Gaussian statistic.
We set $\mathrm{FAP} = 0.05$ and find the value of $\Gamma_\ast$ which satisfies Eq.~\eqref{eq:calcFAP}. To determine the threshold, we use 500 test data of simulated Gaussian noise.

Once we obtain the threshold, we determine the minimum SNR for detection by simulating data with GW signal and setting the detection probability to $\Sub{p}{det} = 0.9$.
For signal injection, the values of SNR and duty cycle are taken from $\rho \in [0.2, 4.0]$ and $\log_{10}\xi \in [-3.0, -1.0]$ with the interval of $\Delta\rho =0.2$ and $\Delta\log_{10}\xi = 0.2$. We prepare 500 data for each injection value, count the number of data satisfying $\Gamma_\ast<\Gamma $, and obtain the detection probability as a function of $\rho$ for each $\xi$. From this, we can find the minimum value of $\rho$ that gives $\Sub{p}{det} = 0.9$.
To evaluate the statistical fluctuation due to the randomness of the signal and the noise, we independently carry out the whole process four times.

Figure~\ref{fig:resnetresult} summarises the results, showing the minimum detectable SNRs for the four methods, i.e., the non-Gaussian statistic based on DF03, the shallower CNN, the deeper CNN, and the residual network. For the range of $-3.0 \leq \log_{10} \xi \leq -2.0$, all deep learning methods show a comparable performance to the non-Gaussian statistic.
For $-1.75 < \log_{10} \xi$, we see that the residual network performs as well as the non-Gaussian statistic, while the performance of the shallower and deeper CNNs gets worse. The deviation between the residual network and the deeper CNN, which have almost the same number of tunable parameters, clearly shows the advantage of using the residual blocks.

\subsection{Computational time}

At the end of this section, we list the computational times of the neural networks and the non-Gaussian statistic. The computational time of the non-Gaussian statistic is defined by the time to carry out the grid search for 500 test data. For the neural networks, we measure the time that the trained models spend analyzing 500 test data. We use CPU Intel(R) Xeon(R) CPU E5-1620 v4 @ 3.50GHz ($224$ GFLOPS) for the non-Gaussian statistic and GPU Quadro GV100 ($16.6$ TFLOPS in single precision) for the neural networks. Table.~\ref{tab: comp speed} shows the comparison of the computational time and the ratio with respect to the non-Gaussian statistic. Note that here we performed a simple grid search to find the maximum value of the non-Gaussian statistics, but the computational time could be improved by applying a fast grid search algorithm. Even considering this point and the difference in the computational power between the CPU and the GPU, deep learning shows a clear advantage in computational time. This can be fruitful when we apply deep learning for longer strain data that is reasonably expected for a realistic situation.

\begin{table}[t]
    \centering
    \caption{Computational times of different methods for the detection problem.}
    \label{tab: comp speed}
    \begin{ruledtabular}
        \begin{tabular}{lll}
            method & time [sec] & ratio \\ \hline
            non-Gaussian statistic & $1.13 \times 10^4$ & 1\\
            shallower CNN & $2.54 \times 10^{-2}$ & $2.25 \times 10^{-6}$\\
            deeper CNN & $7.96 \times 10^{-2}$ & $7.04\times 10^{-6}$\\
            residual network & $7.66 \times 10^{-2}$ & $6.78 \times 10^{-6}$
        \end{tabular}
    \end{ruledtabular}
\end{table}

\section{Estimating duty cycle}
\label{sec: parameter estimation}

In this section, we demonstrate neural network applications for parameter estimation. We take two approaches. First, the neural network is trained to output the estimated values of the duty cycle and the SNR. In the second approach, we treat parameter estimation as a classification problem by dividing the range of duty cycle values into four classes. 
The first method is more straightforward and can directly give the value of $\xi$, while we show that the estimation gets biased when the duty cycle is relatively small ($\xi \lesssim 10^{-3}$). The second approach can predict only the rough range of $\xi$, but it shows reasonable performance even for smaller duty cycle $\xi \sim 10^{-4}$.

\subsection{First approach: direct estimation of the duty cycle and the SNR}
We train the neural network to predict the value of the duty cycle and the SNR. We use the structure of the residual network shown in Table.~\ref{tab: structure of resnet} and Fig.~\ref{fig: our residual block} by removing the softmax layer. The weight update is repeated for $10^5$ times.
The training data is generated by sampling the duty cycle from the log uniform distribution on $[10^{-2}, 10^0]$ and the SNR from the uniform distribution on $[1, 60]$.
To make the training easier, the injection parameters are normalized by
\begin{equation}
    \hat{Q} = \frac{2Q-Q_\mathrm{min} - Q_\mathrm{max}}{Q_\mathrm{max} - Q_\mathrm{min}}\,,
\end{equation}
where $Q=\{\log_{10} \xi, \rho\}$ is the injected values, and $Q_\mathrm{min}$ and $Q_\mathrm{max}$ are the minimum value and the maximum value of the training range, respectively. By this normalization, $\hat{Q}$ has the range $[-1,1]$. The outputs of the neural network directly correspond to the estimated values of $\hat{Q}$. 
We use the L1 loss (Eq.~\eqref{eq:L1loss}) as the loss function. We set the batch size to 512. The update algorithm is Adam with the learning rate of $10^{-5}$.

\begin{figure}[t]
    \centering
    \includegraphics[width=8.5cm]{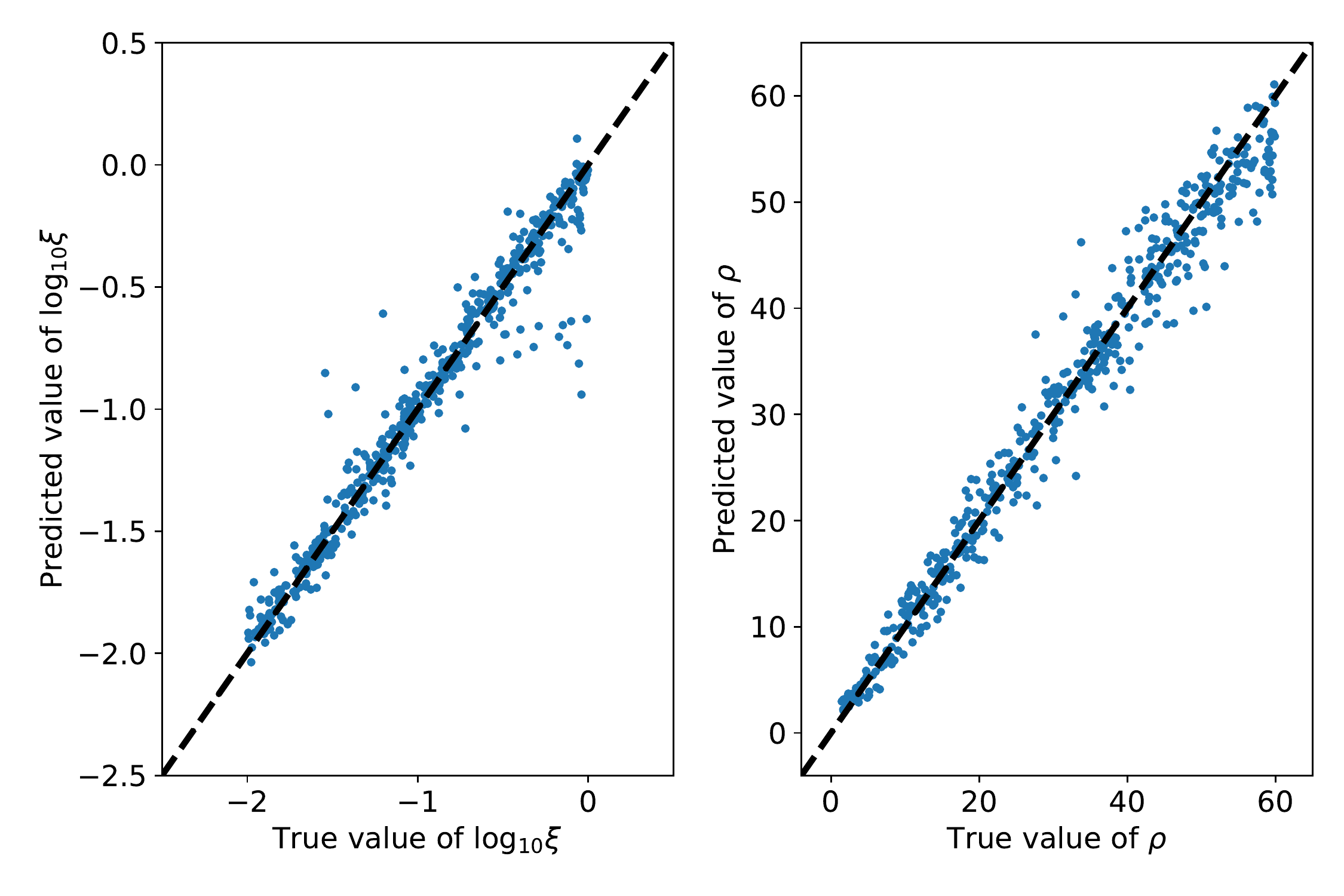}
    \caption{Parameter estimation of the duty cycle (left) and the SNR (right) by the neural network. The scatter plot shows the true value on the horizontal axis and the predicted value on the vertical axis. The diagonal line represents equal values for predicted and true values.
    }\label{fig:true pred diagram}
\end{figure}

We test the trained residual network with the newly-generated data with the parameters sampled from the same distributions as one of the training data. Figure~\ref{fig:true pred diagram} is the scatter plot comparing the true values with the predicted values. We can see that the neural network can recover the true values reasonably well.

In order to evaluate the performance quantitatively, let us define the average and standard deviations of the error as 
\begin{align}
    \overline{\delta Q} &\coloneqq \frac{1}{N}\sum_{n=1}^N \left(Q^{\mathrm{pred}}_n - Q^{\mathrm{true}}_n\right)\,,\\
    \sigma[\delta Q] &\coloneqq \sqrt{\frac{1}{N}\sum_{n=1}^N \left(Q^{\mathrm{pred}}_n - Q^{\mathrm{true}}_n\right)^2}\,,
\end{align}
where $N$ is the number of the test data, $Q^\mathrm{pred}_n$ and $Q^\mathrm{true}_n$ are respectively the predicted value and the true value of the quantity $Q=\{\log_{10}\xi, \rho\}$ of the $n$ th test data.
Table.~\ref{tab: parameter estimation result} shows $\overline{\delta Q}$ and $\sigma[\delta Q]$ obtained by using 500 test data. 
The duty cycle and SNR are randomly sampled from a uniform distribution on $\log_{10}\xi \in [-2, 0]$ and $\rho \in [1, 60]$.
For both the duty cycle and the SNR, $\overline{\delta Q}$ is much smaller than $\sigma[\delta Q]$. 
From this, we can conclude that the neural network predicts the duty cycle and the SNR without bias.
\begin{table}[t]
    \centering
    \caption{Averages and standard deviations of errors in $\log_{10}\xi$ and $\rho$.}
    \label{tab: parameter estimation result}
    \begin{ruledtabular}
        \begin{tabular}{llll}
            $\overline{\delta\log_{10}\xi}$ & $\sigma[\delta\log_{10}\xi]$ & $\overline{\delta\rho}$ & $\sigma[\delta\rho]$ \\ \hline
            $-1.29\times 10^{-5}$ & 0.11 & $-8.90\times 10^{-2}$ & 2.97  
        \end{tabular}
    \end{ruledtabular}
\end{table}

\begin{figure*}[t]
    \begin{tabular}{cc}
        \begin{minipage}[t]{0.48\hsize}
        \centering
        \includegraphics[width=0.95\hsize]{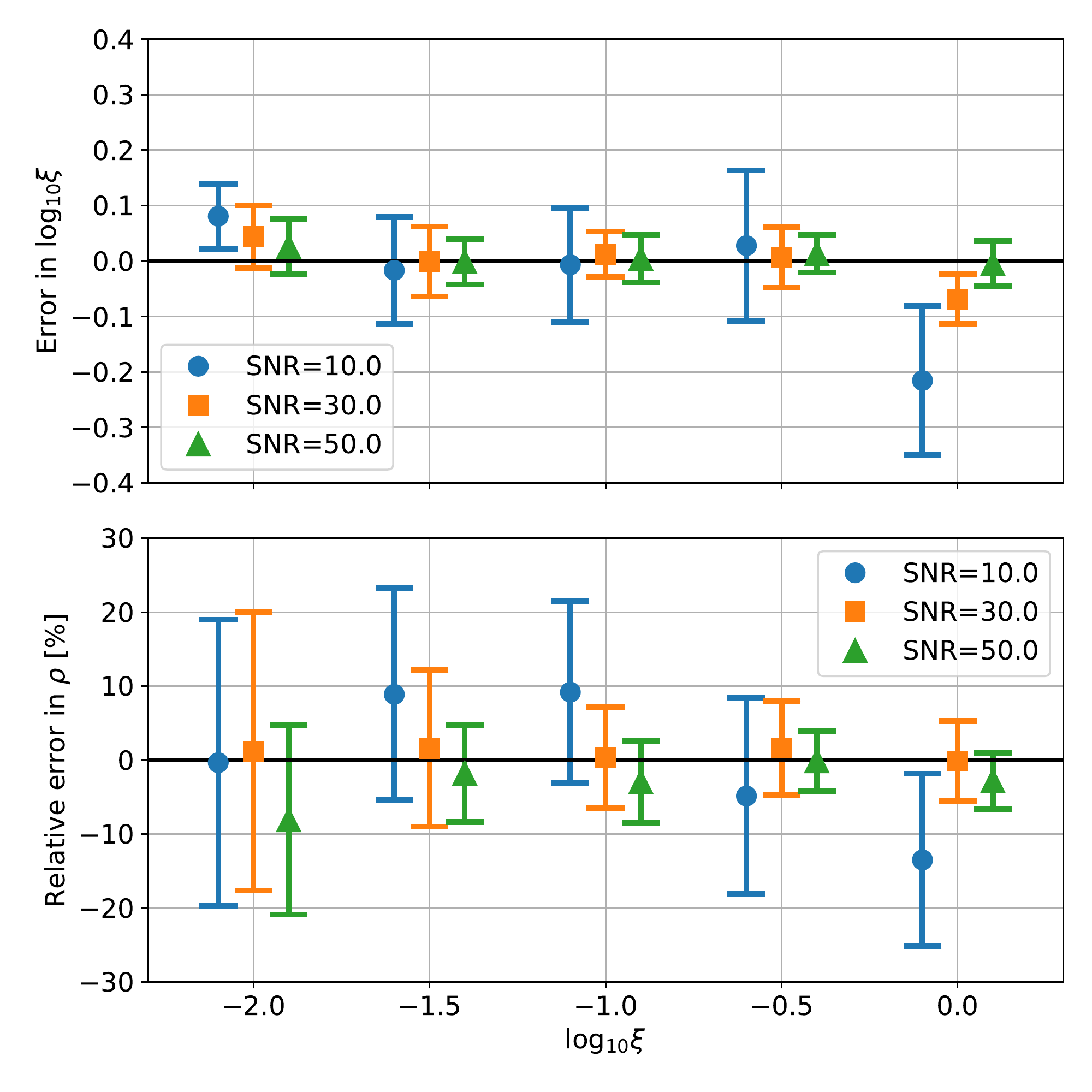}
        \end{minipage} &
        \begin{minipage}[t]{0.48\hsize}
        \centering
        \includegraphics[width=0.95\hsize]{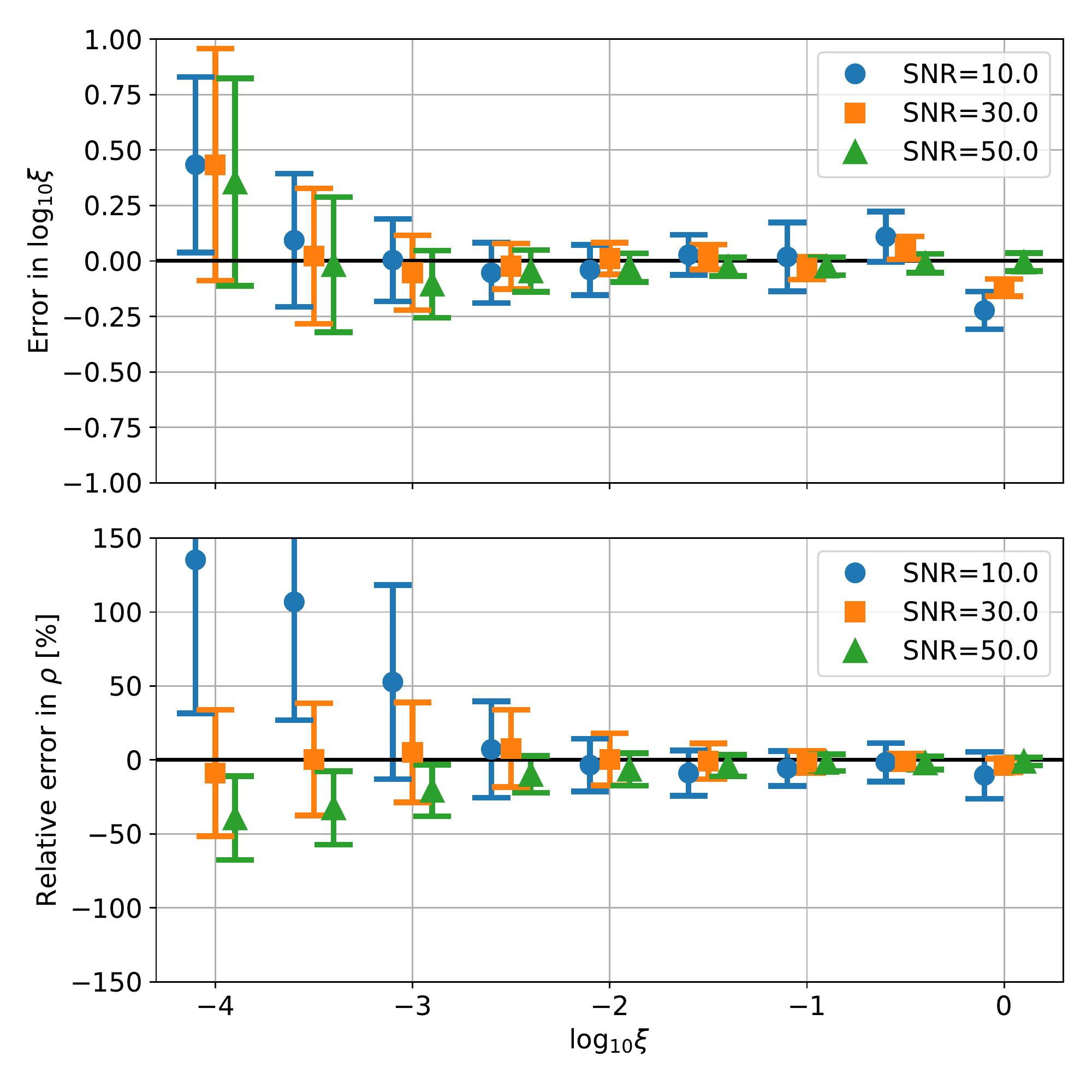}
        \end{minipage}
    \end{tabular}
    \caption{
    Errors in the duty cycle (top) and the SNR (bottom) for different fiducial values of the duty cycle. The blue circles, orange squares, and green triangles respectively show the results of the data sets with the fiducial SNR of 10, 30, and 50. Each dot shows the average of the error, and the error bar represents the standard deviation of the errors. The left panels show the results of the neural network trained with $\log_{10} \xi \in [-2,0]$, and the right panels are the ones trained with $\log_{10} \xi \in [-4,0]$.}
    \label{fig:error vs logxi}
\end{figure*}%

To further check the performance in detail, in the left panel of Fig.~\ref{fig:error vs logxi}, we plot the average errors of $\log_{10}\xi$ (top) and $\rho$ (bottom) for different fiducial parameter values. The error bars indicate their standard deviations. To make this plot, we sample $\log_{10}\xi$ from $-2$ to $0$ by the interval of $0.5$. For each duty cycle, we prepare datasets with SNR $10$, $30$, and $50$. Each dataset contains $500$ realizations. Note that we do not use the relative error for $\log_{10}\xi$ because the target value can be close to zero, which causes divergence in the relative error.
First, we find from both left panels that the error variance reasonably increases as the SNR decreases. The estimation of the duty cycle and SNR seems not to be biased except when the fiducial value is at the border of the training range ($\log_{10} \xi = 0$ and $-2$), and the SNR is small ($\rho=10$).

In the right panels of Fig.~\ref{fig:error vs logxi}, we show the results in which we include lower values of the duty cycle for the training, $\log_{10}\xi \in [-4,0]$. We find that the error variances of both the duty cycle and the SNR significantly increase as the duty cycle decreases for $\log_{10} \xi \lesssim -3$. For the duty cycle, the systematic bias is smaller than the variance. On the other hand, for the SNR, we find a clear bias that the neural network tends to output a larger SNR than the true value when $\rho=10$ and a smaller SNR when $\rho=50$. We find from test runs that such biases tend to increase when we use a shorter data length. From this, we can infer that the bias arises because the data length is too short. In fact, with the data length of $N=10^4$ used throughout this paper, the burst can be absent in the strain data for $\xi \sim 10^{-4}$.

\subsection{Second approach: Classification problem}
As a second approach, we consider the classification problem. we divide the range of duty cycle values into four categories and assign the class index as the following
\begin{equation}
    \textrm{class index} = 
    \begin{cases}
        1 & (-1 \leq \log_{10} \xi < 0) \\
        2 & (-2 \leq \log_{10} \xi < -1) \\
        3 & (-3 \leq \log_{10} \xi < -2) \\
        4 & (-4 \leq \log_{10} \xi < -3)\,.
    \end{cases}
\end{equation}
Again, we use the residual network with the structure shown in Table.~\ref{tab: structure of resnet} and Fig.~\ref{fig: our residual block}, but the last fully connected layer and the softmax layer are modified to have four-dimensional outputs.

The training procedure is as follows. The weight update is repeated for $10^5$ times. The input data are normalized in the same way as the detection problem (see Eq.~\eqref{eq: normalized strain}).
The duty cycle is sampled from the log uniform distribution on $[10^{-4}, 10^0]$, and the SNR is sampled from the uniform distribution on $[1, 60]$.
The batch size is 512, and the update algorithm is Adam with the learning rate of $10^{-5}$. For the loss function, we use the cross-entropy loss Eq. \eqref{eq: cross entropy discrete prob} with $N_\mathrm{class}=4$.

\begin{figure}[t]
    \centering
    \includegraphics[width=0.95\hsize]{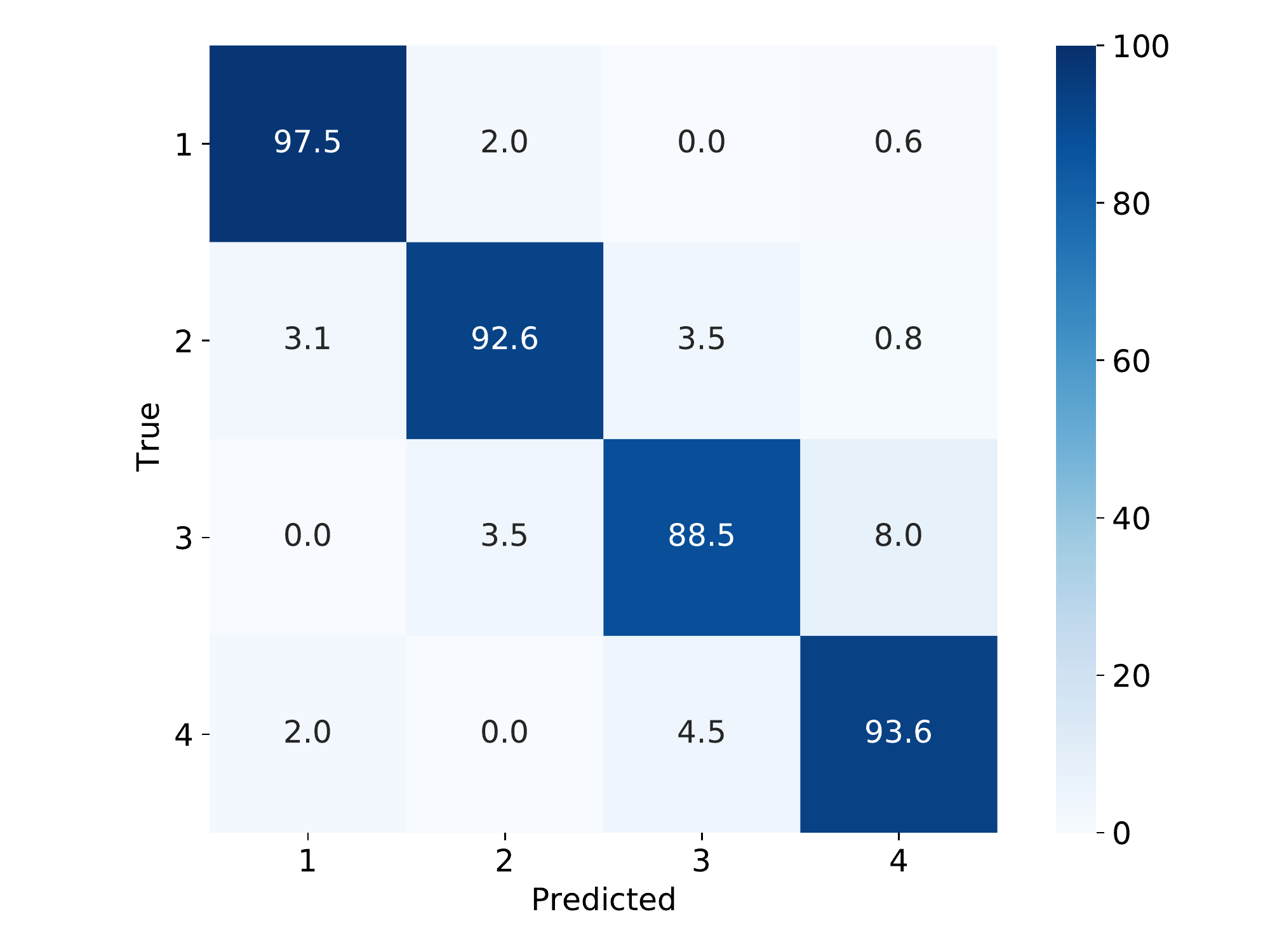}
    \caption{Confusion matrix for the duty cycle estimation. The row and column represent the true label and predicted label, respectively. Each class is labeled by the integer $\{1,2,3,4\}$ and they correspond to $\log_{10} \xi \in [-1,0)$, $[-2,-1)$, $[-3,-2)$, $[-4,-3)$, respectively. The numbers are in the unit of percent and represent the fraction of 
    data classified from the true label to the predicted label.}
    \label{fig: confusion matrix}
\end{figure}

The trained neural network is tested with four datasets; each consists of 512 data and corresponds to the different classes. In the same way as the training data, SNRs are uniformly sampled from the range [1, 60] for all test datasets. 
Figure~\ref{fig: confusion matrix} presents the confusion matrix of the classification by the residual network. We find that 93.1\% of test data are successfully classified to the correct class on average. Also, unlike the direct parameter estimation shown in the previous subsection, we can see that the residual neural network works well even for small values of the duty cycle $\log_{10} \xi \lesssim -2$. Thus, this method could be useful for giving an order of magnitude estimation of the duty cycle.

\begin{figure}[t]
    \centering
    \includegraphics[width=0.95\hsize]{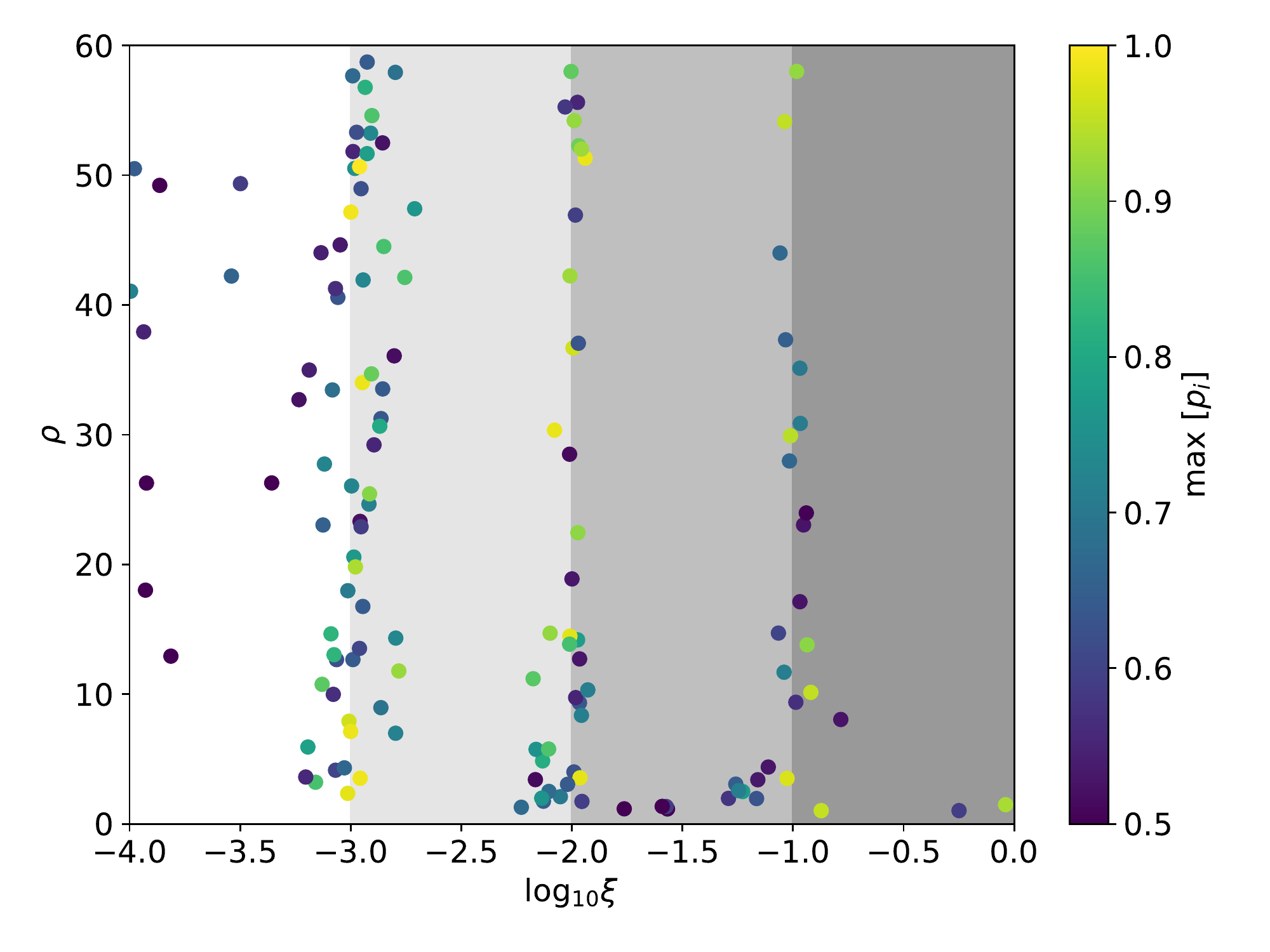}
    \caption{Distribution of the true values of ($\log_{10}\xi, \rho)$ of the misclassified events.
    The color of the dots represents the maximum value of the probability among $\{p_i\}_{i=1,2,3,4}$.}
    \label{fig: misclassified event}
\end{figure}

\begin{figure}[t]
    \centering
    \includegraphics[width=0.95\hsize]{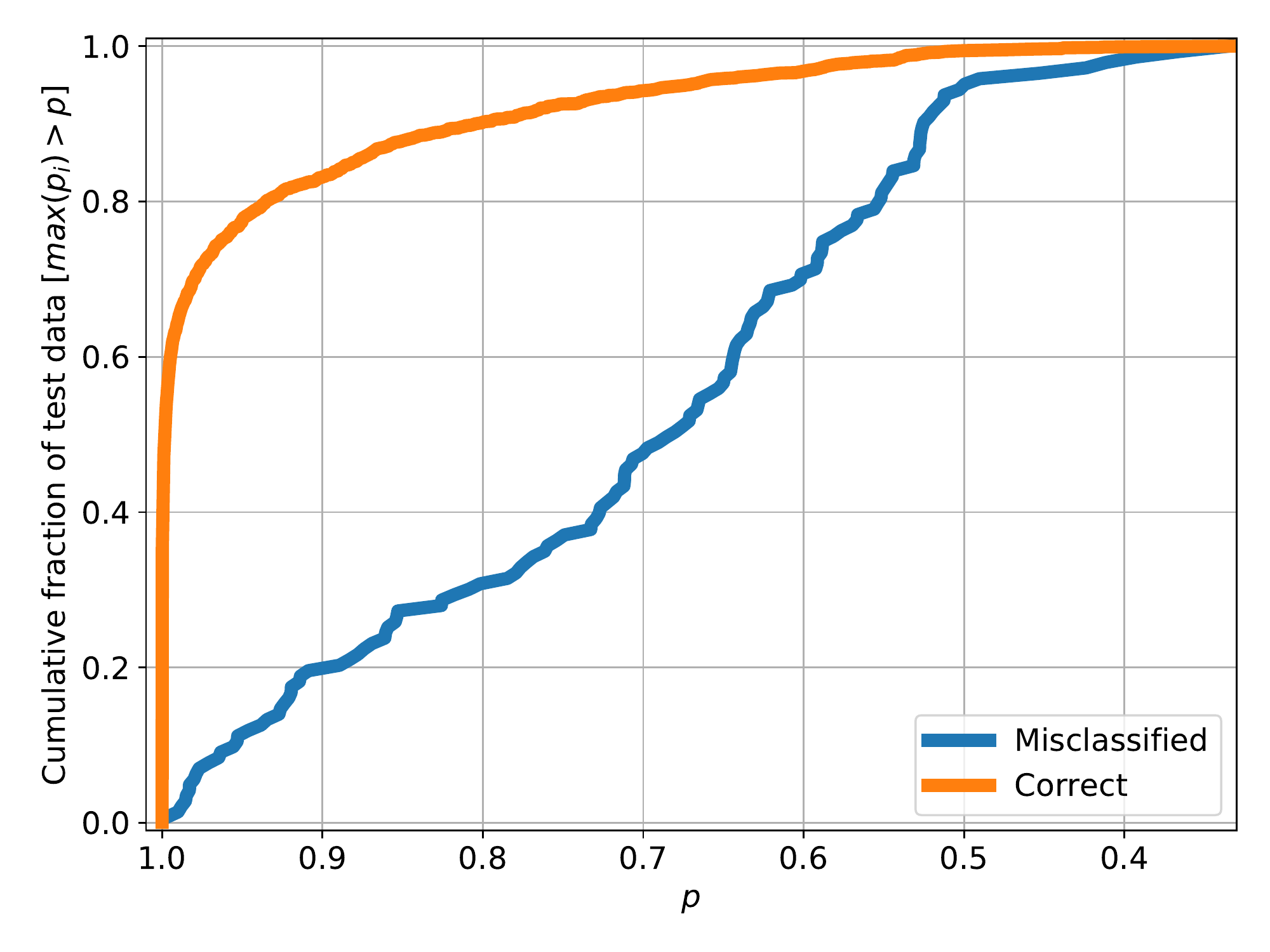}
    \caption{Cumulative fraction of the maximum value of the predicted probabilities in descending order. Blue and orange lines correspond to the misclassified and correctly classified events, respectively. Note that the numbers of correctly classified and misclassified events are different: 1905 events are correctly classified, and 143 events are misclassified.}
    \label{fig: hist max prob}
\end{figure}

Now, we further investigate the misclassified cases. Figure~\ref{fig: misclassified event} shows the scatter plot of misclassified events in the ($\log_{10}\xi$, $\rho$) plane. It clearly shows that the duty cycles of the misclassified events are located at the boundary of the neighboring classes. 
As for the SNR distribution, we find that it is almost uniform, but as expected, there is a tendency that misclassification occurs more for $\rho \lesssim 5$. 
The colors of the dots in the scatter plot represent the maximum values of $p_i$ (the probability of each class), which indicates how confidently the neural network predicts the class. We can see that most of the misclassified events are given with low confidence. 

Figure~\ref{fig: hist max prob} shows the cumulative histograms (cumulating in the reverse direction of $p_i$) for correctly classified events and misclassified events. We can see a clear difference between them. For most of the correctly classified events, the probability of close to 1 is assigned.
On the other hand, we can again see that misclassified events tend to have low confidence. However, $20\%$ of the events are misclassified with $\max~[p_i] > 0.9$. As seen from Fig.~\ref{fig: misclassified event}, they are at the boundary of the neighboring classes, and this would be unavoidable with the classification problem method.

\section{Conclusion}
\label{sec: conclusion}

In this work, we studied applications of convolutional neural networks to the detection and parameter estimation of non-Gaussian stochastic GWs. As for the detection problem, we compared three different configurations of neural networks; shallower CNN, deeper CNN, and residual network. We found that the residual network can achieve comparable sensitivity to the maximum likelihood statistics. We also showed that neural networks have an advantage in computational time compared to the non-Gaussian statistic.

Next, we investigated the estimation of the duty cycle by a neural network with two different approaches. In the first approach, we trained the residual neural network to directly estimate the values of the duty cycle and SNR. We found that the estimation error in $\log_{10}\xi$ is about $\lesssim 0.2$. As for SNR, the neural network can estimate with the relative error of $10 - 20\%$. We found that the estimation of the duty cycle gets biased when we include a small duty cycle for the training $\log_{10}\xi \in [-4,0]$. This could be explained by the shortness of the data length used in this paper.
In the second approach, the parameter estimation was reduced to the classification problem in which the neural network classifies the data depending on the duty cycle. The parameter range was $\log_{10}\xi \in [-4,0]$, and it was divided into four classes with the band of $\Delta\log_{10}\xi = 1$. The neural network could classify the data with an accuracy of 93\% on average.

The present work is the first attempt to apply deep learning to the astrophysical GW background. In this work, we employed the toy model that is used in DF03 where various realistic effects, such as the detector's configuration, noise properties, and waveform model of the bursts are neglected. In particular, detection of the astrophysical GW background would become challenging in the presence of glitch noises and the correlated magnetic noise from Schumann resonances. 
Further study of their effects will be extremely important for applying our method to real data. We leave it as future work with an expectation that deep learning has a high potential to distinguish such troublesome noises from the signal.

\begin{acknowledgements}
This work is supported by JSPS KAKENHI Grant no. JP20H01899. TSY is supported by JSPS KAKENHI Grant no. JP17H06358 (and also JP17H06357), \textit{A01: Testing gravity theories using gravitational waves}, as a part of the innovative research area, ``Gravitational wave physics and astronomy: Genesis''. SK acknowledges support from the research project PGC2018-094773-B-C32, and the Spanish Research Agency through the Grant IFT Centro de Excelencia Severo Ochoa No CEX2020-001007-S, funded by MCIN/AEI/10.13039/501100011033.
SK is supported by the Spanish Atracci\'on de Talento contract no. 2019-T1/TIC-13177 granted by Comunidad de Madrid, the I+D grant PID2020-118159GA-C42 of the Spanish Ministry of Science and Innovation, the i-LINK 2021 grant LINKA20416 of CSIC, and JSPS KAKENHI Grant no. JP20H05853. This work was also supported in part by the Ministry of Science and Technology (MOST) of Taiwan, R.O.C., under Grants No. MOST 110-2123-M-007-002 and 110-2112-M-032 -007- (G.C.L.)
The authors thank Joseph D. Romano for very useful comments. 
\end{acknowledgements}

\appendix
\section{Review of non-Gaussian statistic}
\label{appendix:NG}

Here, we review the properties of the non-Gaussian statistic~\eqref{eq: def nonGaussianLambda}. DF03 compared the non-Gaussian statistic with the standard cross-correlation statistic that is defined by
\begin{equation}
    \Sub{\Lambda}{CC}(h) \coloneqq \frac{\hat{\alpha}^2}{\bar{\sigma}_1 \bar{\sigma}_2}\,,
    \label{eq: def cross correlation statistic}
\end{equation}
where
\begin{equation}
    \hat{\alpha}^2 \coloneqq \bar{\alpha}^2 \Theta(\bar{\alpha}^2)\,, \qquad
    \bar{\alpha}^2 \coloneqq \frac{1}{N}\sum_{k=1}^{N} h_1^k h_2^k\,,
\end{equation}
and $\Theta(x)$ is the Heaviside step function defined by
\begin{equation}
    \Theta(x) = \begin{cases}
    1 & \text{if}\ x\geq 0 \\
    0 & \text{if}\ x < 0\,.
    \end{cases}
\end{equation}
This is obtained by assuming the Gaussian signal model, i.e., $\xi=1$ in Eq.~\eqref{eq: signal model}.

\begin{figure}[t]
    \includegraphics[width=0.9\hsize]{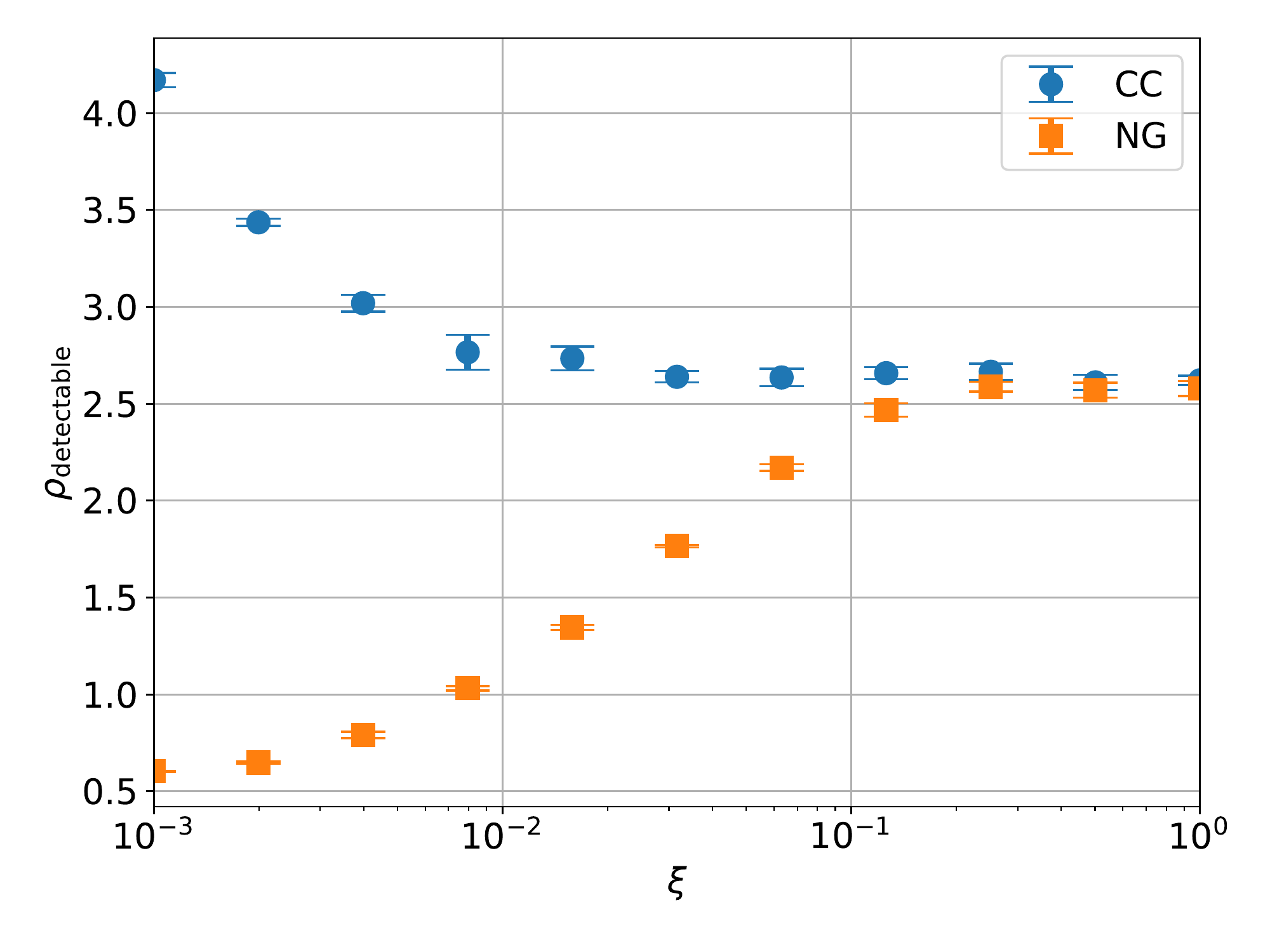}
    \caption{\label{fig:detectablesnr}
    Minimum detectable SNR as a function of the duty cycle $\xi$. Both the false alarm probability and the false dismissal probability are set to be 0.1. Error bars are obtained by four independent runs. The time-series data has the length of $N=10^4$, and the detector's noise variances are $\sigma_1^2 = \sigma_2^2 = 1$. Note that $\rho_\mathrm{90\%}$ represents the minimum detectable SNR with $90\%$ detection probability and is different from that of $\Omega_\mathrm{detectable}$ in Fig.1 of DF03.}
\end{figure}

Here we aim to reproduce the results of DF03 and demonstrate the performance of Eq.~\eqref{eq: def nonGaussianLambda} by simulating time series strain data with the length $N=10^4$.
The maximization of $\SupSub{\lambda}{NG}{ML}$ in Eq.~\eqref{eq: def nonGaussianLambda} requires to explore the parameter space. Here, by following DF03, we substitute the injected values into $\SupSub{\lambda}{NG}{ML}$ instead of maximizing the model parameters. Note that we perform the parameter search to simulate the non-Gaussian statistic for the comparison purpose in the main part of the paper, but the general behavior does not change.

Figure~\ref{fig:detectablesnr} compares the minimum detectable SNR for the standard cross-correlation statistic and the non-Gaussian statistic.
For $\xi>0.1$, their performances are comparable. This can be interpreted that the non-Gaussianity of the signal is not much strong, and taking into account non-Gaussianity does not give a significant advantage. On the other hand, for $\xi < 0.1$, the non-Gaussian statistic outperforms the cross-correlation statistic. It is reasonable because the non-Gaussian statistic is developed based on the same signal model as the one we used for simulating strain data.

\begin{figure}[t]
    \centering
  \includegraphics[width=0.95\hsize]{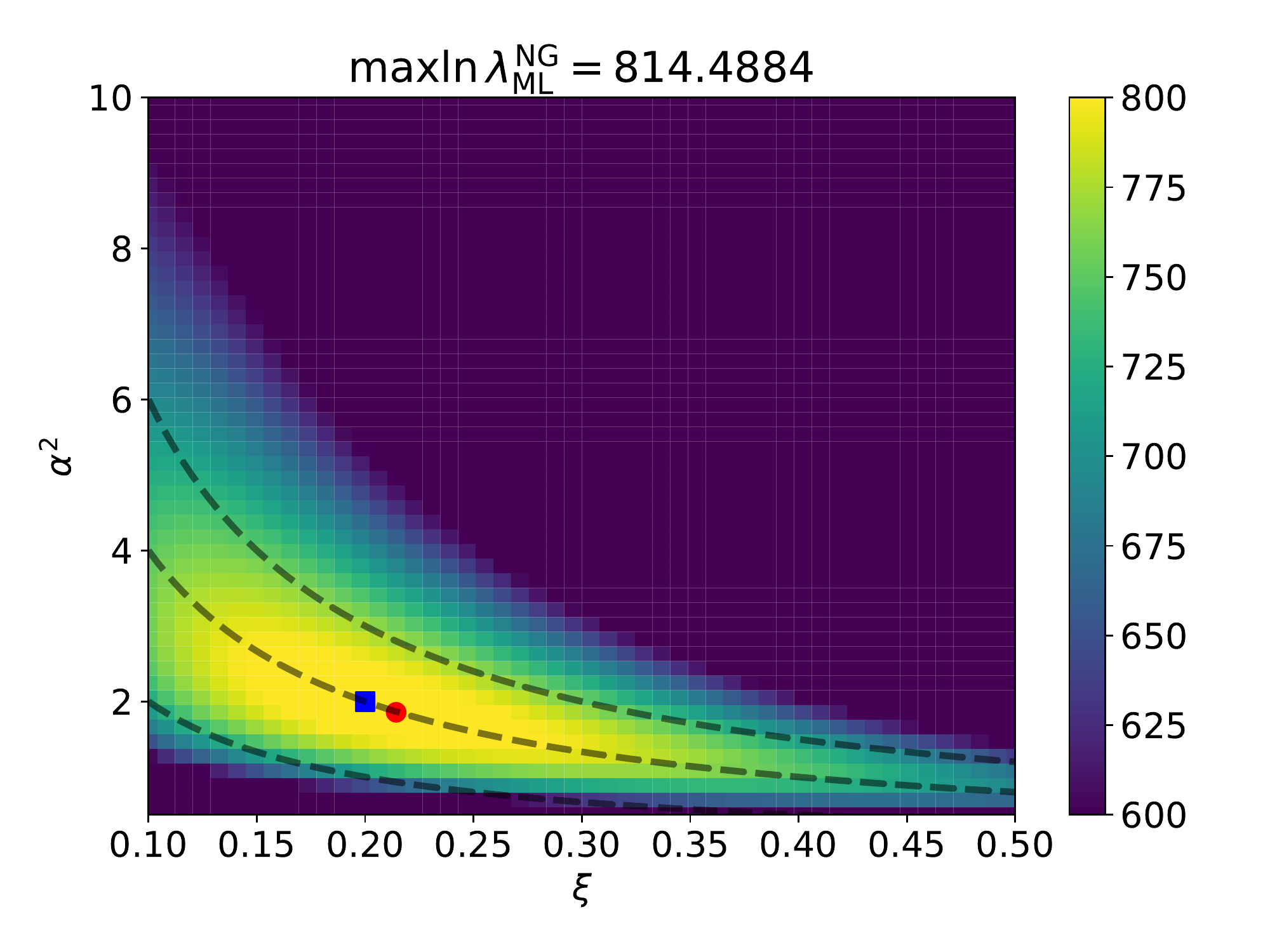}
  \caption{\label{fig:exampleLambdaNG}
  The color map shows the logarithm of $\SupSub{\lambda}{NG}{ML}(\alpha^2, \xi)$. The data length is $N=10^4$, and we set the signal parameters to $\xi=0.2$ and $\rho=40$ (indicated by the blue square). The variances of the detector noises are $\sigma_1^2 = \sigma_2^2 = 1$. The red circle indicates the parameter values of the maximum log-likelihood. Black dashed lines indicate the constant SNR for $\rho = 60,40,20$ from top to bottom.
  It is clearly seen that the likelihood estimator has a degeneracy along the parameter combination that gives the same SNR value.}
\end{figure}

Next, parameter estimation is tested. Figure~\ref{fig:exampleLambdaNG} shows an example of the distribution of the logarithm of $\SupSub{\lambda}{NG}{ML}$ in the $\xi - \alpha^2$ plane. We injected a stochastic signal with $\xi=0.2$ and $\rho = 40$. It is clearly seen that the duty cycle $\xi$ and the amplitude variance $\alpha^2$ of each burst degenerate. We also draw three dashed lines corresponding to different SNRs, $\rho=$20, 40, and 60.
We clearly see that the strong degeneracy exists along the line of constant SNR. In other words, the non-Gaussian statistic is sensitive to the difference in SNR.

\bibliography{references}

\end{document}